\documentclass[aps,pre,amsmath,amssymb,reprint,groupedaddress,showpacs]{revtex4-1}

\usepackage{graphicx}% Include figure files
\usepackage{dcolumn}% Align table columns on decimal point
\usepackage{bm}% bold math

\begin{document}

% Use the \preprint command to place your local institutional report
% number in the upper righthand corner of the title page in preprint mode.
% Multiple \preprint commands are allowed.
% Use the 'preprintnumbers' class option to override journal defaults
% to display numbers if necessary
%\preprint{}

%Title of paper
\title{Electromagnetic space-time crystals. III. Dispersion relations for partial solutions}

% repeat the \author .. \affiliation  etc. as needed
% \email, \thanks, \homepage, \altaffiliation all apply to the current
% author. Explanatory text should go in the []'s, actual e-mail
% address or url should go in the {}'s for \email and \homepage.
% Please use the appropriate macro foreach each type of information

% \affiliation command applies to all authors since the last
% \affiliation command. The \affiliation command should follow the
% other information
% \affiliation can be followed by \email, \homepage, \thanks as well.
\author{G. N. Borzdov}
%\email[]{Your e-mail address}
\email[]{BorzdovG@bsu.by}
%\homepage[]{Your web page}
%\thanks{}
%\altaffiliation{}
\affiliation{Department of Theoretical Physics and Astrophysics, Belarus State University,
Nezavisimosti avenue 4, 220030 Minsk, Belarus}

%Collaboration name if desired (requires use of superscriptaddress
%option in \documentclass). \noaffiliation is required (may also be
%used with the \author command).
%\collaboration can be followed by \email, \homepage, \thanks as well.
%\collaboration{}
%\noaffiliation

%\date{\today}

\begin{abstract}
% insert abstract here
Partial solutions of the Dirac equation describing an electron motion in electromagnetic crystals created by plane waves with linear and circular polarizations are treated. It is shown that the electromagnetic crystal formed by circularly polarized waves possesses the spin birefringence.
\end{abstract}

% insert suggested PACS numbers in braces on next line
% insert suggested keywords - APS authors don't need to do this
%\keywords{}

%\maketitle must follow title, authors, abstract, \pacs, and \keywords
\maketitle

% body of paper here - Use proper section commands
% References should be done using the \cite, \ref, and \label commands
\section{INTRODUCTION}
% Put \label in argument of \section for cross-referencing
%\section{\label{}}
In the band theory of solids, the substitution of the Bloch function into the single-electron steady-state Shr\"{o}dinger equation, due to the periodic nature of a crystal lattice, results in an infinite system of linear homogeneous equations relating scalar Fourier amplitudes of this wave function~\cite{Ani,Davydov}. The existence condition of nontrivial solutions of the system provides the dispersion relation $E=E(\textbf{k})$ defining the dependence of the electron energy $E$ on the wavevector $\textbf{k}$ in an energy band. Because the system is infinite, the dispersion relation is derived by a method of succesive approximations~\cite{Ani,Davydov}. A similar approach is applied in~\cite{gaps} to find the band structure of the space-time crystal created by a standing plane harmonic electromagnetic wave.

In this series of papers~\cite{ESTCp1,ESTCp2}, we treat the electromagnetic space-time crystals (ESTCs) created by three standing plane harmonic waves with mutually orthogonal phase planes and the same frequency. The Dirac equation describing the motion of an electron in ESTC, that is in an electromagnetic field with four-dimensional (4d) periodicity, reduces to an infinite system of linear matrix equations. Each equation relates 13 Fourier amplitudes [bispinors $c(n+s)$] of the wave function, where the multi-index $n=(n_1,n_2,n_3,n_4)$ is a point of the integer lattice $\mathcal{L}$ with even values of the sum $n_1+n_2+n_3+n_4$, and $s=(s_1,s_2,s_3,s_4)\in S_{13}\subset\mathcal{L}$ takes all 13 values $S_{13}$ satisfying the condition $g_{4d}(s)\equiv \max\{|s_1|+|s_2|+|s_3|,|s_4|\}=0, 1$. The relatively simple structure of equations has made it possible to obtain the fundamental solution of this infinite system in the form of the projection operator $S$ defining the subspace of solutions in the infinite dimensional linear space $V_C$ of multi-spinors $C =\{c(n),n \in {\mathcal L}\}$~\cite{ESTCp1}. For any $C_0\in V_C$, $C=\mathcal{S}C_0$ specifies an exact partial solution, i.e., the bispinor wave function with the set of Fourier amplitudes $\{c(n), n \in \mathcal{L}\}=\mathcal{S}C_0$ satisfies the Dirac equation for the problem under consideration. The fundamental solution is obtained by a recurrent process. It is expressed in terms of an infinite series of projection operators. At each step of the recurrent process, the relations presented in~\cite{ESTCp1,ESTCp2} provide the exact fundamental solution to an infinite set of independent finite systems of interrelated equations [fractal clusters of equations (FCE)]. It can be described as a 4d lattice of such clusters. The aggregation scheme for FCE, presented in~\cite{ESTCp2}, is devised to simplify computations and to minimize volumes of data files at calculating the corresponding projection operators. It makes possible to expand FCEs to finite models of ESTC~\cite{ESTCp2} of any desired size and to obtain families of approximate partial solutions of the Dirac equation. To compare in accuracy various approximate solutions of the Dirac equation, obtained in the framework of these models, we use the criterion suggested in~\cite{ESTCp2}, i.e., the relative residual $\mathcal{R}$ at the substitution of an approximate solution into the Dirac equation. The way in which its application reveals, in particular, dispersion relations is illustrated in this paper on several examples. We apply the general technique developed in the preceding papers~\cite{ESTCp1,ESTCp2} to two types of electromagnetic space-time crystals, denoted ESTC1 and ESTC2, formed by plane waves with linear and circular polarizations, respectively.
In section~\ref{sec:free}, we discuss in detail the interrelation between the free space solutions of the Dirac equation and the approximate solutions which describe an electron in ESTC at limiting process to the vanishing field. ESTC1 and ESTC2 are treated in sections~\ref{sec:ESTC1} and~\ref{sec:ESTC2}, respectively. For the problem under consideration, the Dirac equation reduces to an infinite system of matrix equations, where the interconnections between equations are defined~\cite{ESTCp1} by 12 matrix functions and 56 scalar coefficients. Appendix gives the expressions for them in an explicit form.

\section{\label{sec:free}Free space solution as a limit case}
In the absence of electromagnetic field, the Dirac equation for the wave function $\Psi(\bm{x})=\exp(i \bm{K}\cdot\bm{x})c_0$ reduces to the form [see Eq.~(37) in~\cite{ESTCp2} with $n=n_o=(0,0,0,0)$]:
\begin{equation}\label{P0c0}
    P_0c_0=0,\quad P_0=\frac12 U - \frac{1}{2q_4} \left(\alpha_4+\sum_{k=1}^3 q_k\alpha_k\right).
\end{equation}
Here, $\Psi$ and $c_0$ are the bispinors, $\bm{x}=(\textbf{r},i c t)$, $\bm{K} = (\textbf{k},i \omega /c )$ is the four-dimensional wave vector,  $U$ is the unit $4\times 4$ matrix, $\alpha_j$ are the Dirac matrices, and we use dimensionless parameters
\begin{eqnarray}
% \nonumber to remove numbering (before each equation)
  \bm Q = (\textbf{q},i q_4)&=&{\bm K}/\kappa_e,\nonumber\\
  \textbf{q} = q_1\textbf{e}_1+q_2\textbf{e}_2+q_3\textbf{e}_3&=&\frac{\hbar \textbf{ k}}{m_e c}, \quad   q_4 = \frac{\hbar \omega}{m_e c^2},\label{Qq}
\end{eqnarray}
where $\kappa_e=m_e c/ \hbar$, $c$ is the speed of light in vacuum, $\hbar$ is the Planck constant, $e$ is the electron charge, $m_e$ is the electron rest mass. The existence condition $|P_0|=0$ of a nontrivial solution $c_0\neq 0$ results in the dispersion relation
\begin{equation}\label{q42}
    q_4^2=1+\textbf{q}^2,
\end{equation}
which, in terms of the energy $E=m_{e}c^2q_4 =\hbar\omega$ and the momentum $\textbf{p}=m_{e}c\textbf{q}=\hbar\textbf{k}$, takes the familiar form~\cite{Novo}
\begin{equation}\label{E2}
    E^2=c^2(m_{e}^2c^2+\textbf{p}^2).
\end{equation}
Once this condition is satisfied, i.e.,
\begin{equation}\label{q4pm}
    q_4=\pm q_{40},\quad q_{40}=\sqrt{1+\textbf{q}^2},
\end{equation}
Eq.~(\ref{P0c0}) splits into two independent equations for positive ($q_4=q_{40}$) and negative ($q_4=-q_{40}$) frequency domains
\begin{equation}\label{Pcpm}
    P_{-}c_{+}=0,\quad P_{+}c_{-}=0,
\end{equation}
where
\begin{equation}\label{Ppm}
    P_{\pm}=\frac12 U \pm\frac{1}{2q_{40}} \left(\alpha_4+\sum_{k=1}^3 q_k\alpha_k\right)
\end{equation}
are the Hermitian projection matrices specifying the two-dimensional subspaces of solutions at these domains ($c_{\pm}=P_{\pm}c_0$ for any $c_0$) and satisfying the relations
\begin{eqnarray}
% \nonumber to remove numbering (before each equation)
  P_{\pm}^{\dag}=P_{\pm}^2=P_{\pm}, \quad P_{\pm}P_{\mp}=0,\nonumber\\
  P_{+}+P_{-}=U, \quad tr(P_{\pm}) = 2. \label{Pmprop}
\end{eqnarray}

It should be emphasized that, in the case of a nonvanishing field, all projection operators $\rho_0(n)=P(n)$ (see Eq.~(16) in~\cite{ESTCp1}) have the trace $tr[P(n)]=4$, and the fundamental solution $\mathcal{S}$ is obtained in~\cite{ESTCp1} without recourse to any dispersion relation. To explain the interrelation between the two problems, it is sufficient to assume that the potential of the electromagnetic field is small and to use the most simple finite model of ESTC, $0$-model described in~\cite{ESTCp2}. In the frame of this model, we obtain the following relations:
\begin{equation}\label{Sprime}
    \mathcal{S'}=\mathcal{U} - \rho_0(n_o),
\end{equation}
\begin{eqnarray}
% \nonumber to remove numbering (before each equation)
  S(n)&=&U\delta(n-n_o)-R_0(n,n_o,n_o),\nonumber\\
  n&=&\{n_1,n_2,n_3,n_4\}\in S_{13}, \label{Sn}
\end{eqnarray}
\begin{equation}\label{psix}
    \Psi(\bm{x})=\sum_{n\in S_{13}}c(n)e^{i \varphi_n(\bm{x})}, \quad c(n)=S(n)c_0,
\end{equation}
\begin{eqnarray}
% \nonumber to remove numbering (before each equation)
  \varphi_n(\bm{x})&=&[\bm{K} + \bm{G}(n)]\cdot \bm{x} \nonumber\\
  &=&(\textbf{k}+k_0\textbf{n})\cdot\textbf{r} - (\omega+\omega_0n_4)t \nonumber\\
  &=&2\pi[(\textbf{n}+\textbf{q}/\Omega)\cdot\textbf{r}' - (n_4+q_4/\Omega)X_4], \label{phin}\\
  \Omega&=&\frac{\hbar \omega_0}{m_{\rm e} c^2}, \nonumber
\end{eqnarray}
where $\mathcal{S'}$ is the fundamental solution of equation $P(n_o)c_0=0$, $\mathcal{U}$ is the unit operator, $\delta(n-n_o)$ is the Kronecker delta, matrix $R_0(n,n_o,n_o)$ is defined in~\cite{ESTCp1}, $\textbf{n}=n_1\textbf{e}_1+n_2\textbf{e}_2+n_3\textbf{e}_3$, $\omega_0$ is the frequency of electromagnetic field, $k_0=\omega_0/c=2\pi/\lambda_0$ is the wave number, $\textbf{r}'=\textbf{r}/\lambda_0=X_1\textbf{e}_1+X_2\textbf{e}_2+X_3\textbf{e}_3$ and $X_4=ct/\lambda_0$ are the dimensionless coordinates.
The spectral expansion of the matrix $S(n_o)$ has the form
\begin{equation}\label{Sn0}
    S(n_o)=S_{+}P_{+}+S_{-}P_{-},
\end{equation}
where
\begin{equation}\label{Spm}
    S_{\pm}=\frac{I_A}{I_A + (q_4 \mp q_{40})^2}
\end{equation}
are the eigenvalues, $P_{\pm}$ are given by Eq.~(\ref{Ppm}), and the parameter $I_A$ specifies the intensity of the electromagnetic field creating ESTC (see Eq.~(21) in~\cite{ESTCp1}).

Let us consider the family $\Psi(\bm{x},c_0,q_4)$ of functions $\Psi$~(\ref{psix}) at given vector $\textbf{q}$. First, as the initial approximation, called below $0'$--model, we treat its truncated form
\begin{equation}\label{psitr}
    \Psi'(\bm{x},c_0,q_4)=e^{i \bm{K}\cdot\bm{x}}S(n_o)c_0.
\end{equation}
Then Eqs.~(40)--(43) in~\cite{ESTCp2} give
\begin{equation}\label{Rc0q4}
    \mathcal{R}(c_0,q_4)=\sqrt{I_A\frac{c_0^{\dag} S(n_o)c_0}{c_0^{\dag} [S(n_o)]^2c_0}}.
\end{equation}
If $c_0=c_{\pm}$ is an eigenvector of $S(n_o)$, Eq.~(\ref{Rc0q4}) reduces to the relation
\begin{equation}\label{Rpmq4}
    \mathcal{R}_{\pm}(q_4)=\sqrt{I_A+(q_4\mp q_{40})^2}
\end{equation}
which, at $q_4=\pm q_{40}$, gives
\begin{eqnarray}
% \nonumber to remove numbering (before each equation)
  \mathcal{R}_0&=&\mathcal{R}_{+}(q_{40})=\mathcal{R}_{-}(-q_{40})=\sqrt{I_A},\nonumber\\
  \mathcal{R}_{+}(-q_{40})&=&\mathcal{R}_{-}(q_{40})=\sqrt{I_A + 4q_{40}^2}\nonumber\\
   &=&2\sqrt{1+\textbf{q}^2+I_A/4}. \label{R0}
\end{eqnarray}
Thus, in this approximation, the free space solutions $\Psi'(\bm{x},c_{\pm},\pm q_{40})$ provide the minimum value $\mathcal{R}_0$ for the relative residual parameter $\mathcal{R}$. Let now $I_A$ tends to zero. The function $\Psi'$ can be treated as an approximate solution if, and only if $\mathcal{R}(c_0,q_4)\ll 1$, i.e., $c=c_{+}$ and $|q_4-q_{40}|\ll 1$, alternatively, $c=c_{-}$ and $|q_4+q_{40}|\ll 1$. If $|q_4|\neq q_{40}$, one obtains only the trivial solution $S(n_o)=0$ as the limiting case at $I_A\rightarrow 0$. The two physically relevant exact solutions, described by
\begin{eqnarray}
% \nonumber to remove numbering (before each equation)
  S_{+}&=&1,\; S_{-}=0,\; S(n_o)=P_{+},\; \mathcal{R}_{+}(q_{40})=0, \label{posdom}\\
  S_{+}&=&0,\; S_{-}=1,\; S(n_o)=P_{-},\; \mathcal{R}_{-}(-q_{40})=0, \label{negdom}
\end{eqnarray}
arise as limiting cases ($I_A\rightarrow 0)$ of Eqs.~(\ref{Sn0}), (\ref{Spm}) and (\ref{Rpmq4}) at $q_4=q_{40}$ and $q_4=-q_{40}$, respectively.

To analyze the solution $\Psi(\bm x)$ for dependence on the amplitude $c_0$, one can use any basis of the four-dimensional bispinor space. In particular, it is convenient to use the orthonormal basis
\begin{eqnarray}
% \nonumber to remove numbering (before each equation)
  c_1&=&\frac{1}{\delta}\left(
                        \begin{array}{c}
                          1+q_{40} \\
                          0 \\
                          q_3 \\
                          q_1+i q_2 \\
                        \end{array}
                      \right),\quad c_2=\frac{1}{\delta}\left(
                                                          \begin{array}{c}
                                                            0 \\
                                                            1+q_{40} \\
                                                            q_1-i q_2 \\
                                                            -q_3 \\
                                                          \end{array}
                                                        \right),\nonumber\\
  c_3&=&\frac{1}{\delta}\left(
                        \begin{array}{c}
                          q_3 \\
                          q_1+i q_2 \\
                          -1-q_{40} \\
                          0 \\
                        \end{array}
                      \right),\quad c_4=\frac{1}{\delta}\left(
                                                          \begin{array}{c}
                                                            q_1-i q_2 \\
                                                            -q_3 \\
                                                            0\\
                                                            -1-q_{40}\\
                                                          \end{array}
                                                        \right), \label{c1234}
\end{eqnarray}
where $\delta=\sqrt{2q_{40}(1+q_{40})}$, and $P_{\pm}$~(\ref{Ppm}) can be written as
\begin{equation}\label{Ppmcij}
    P_{+}=c_1\otimes c_1^{\dag}+c_2\otimes c_2^{\dag},\quad P_{-}=c_3\otimes c_3^{\dag}+c_4\otimes c_4^{\dag}.
\end{equation}
The notations $c_{+}$ and $c_{-}$ denote below any orthonormal bispinors from the two-dimensional subspaces defined by the projection matrices $P_{+}$ and $P_{-}$, respectively, i.e., $P_{\pm}c_{\pm}=c_{\pm}, c_{\pm}^{\dag}c_{\pm}=1, c_{\pm}^{\dag}c_{\mp}=0$.

Let us now take into account all 13 Fourier  amplitudes $c(n)$ of $\Psi$~(\ref{psix}). As example, we treat here ESTC1 composed of six linearly polarized waves with the amplitudes (see Eq.~(2) in~\cite{ESTCp1})
\begin{eqnarray}
% \nonumber to remove numbering (before each equation)
  \textbf{A}_1&=&-\textbf{A}_4=A_m\textbf{e}_2,\nonumber\\
  \textbf{A}_2&=&-\textbf{A}_5=A_m\textbf{e}_3,\nonumber\\
  \textbf{A}_3&=&-\textbf{A}_6=A_m\textbf{e}_1, \label{A16cr1}
\end{eqnarray}
where $A_m$ is a real scalar amplitude, $I_A=12A_m^2$. In this numerical example, we assume $\Omega=0.1, q_1=q_2=0, q_3=0.02$ [see Eqs.~(\ref{Qq}) and (\ref{phin})]. At given $c_0=c_j$ and $c_0=c_{\pm}$, Eq.~(43) in \cite{ESTCp2} and Eq.~(\ref{psix}) give functions $\mathcal{R}(c_j,q_4)$ and $\mathcal{R}(c_{\pm},q_4)$.

%%%%%%%%%%%%%%%%%%%%%%%%%%%%%figure1%%%%%%%%%%%%%%%
\begin{figure}
\includegraphics{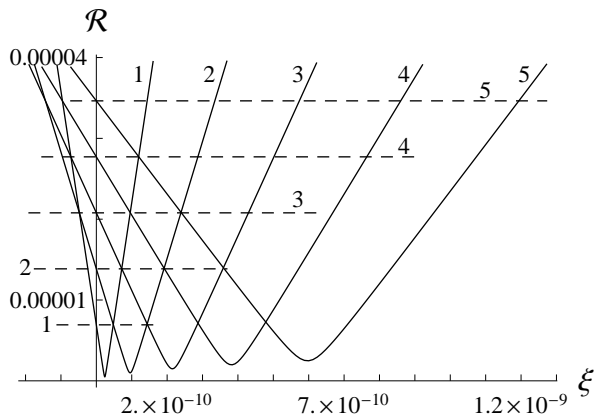}
\caption{\label{p3fig1}The dependence of $\mathcal{R}=\mathcal{R}_{+}(\xi)$ on $\xi=q_4-q_{40}$ for $\Psi$~(\ref{psix}) (solid curves) and $\Psi'$~(\ref{psitr})    (dash curves) at amplitude (1)~$A_m=2\times 10^{-6}$, (2)~$A_m=4\times 10^{-6}$, (3)~$A_m=6\times 10^{-6}$, (4)~$A_m=8\times 10^{-6}$, and (5)~$A_m=10^{-5}$.}
\end{figure}
%%%%%%%%%%%%%%%%%%%%%%%%%%%%%%%%%%%%%%%%%%%%%%%%%%

It follows from the results of our numerical evaluations that in the vicinity of $q_{40}$ (see Fiq.~\ref{p3fig1}) $\mathcal{R}_{+}(q_4)\equiv\mathcal{R}(c_1,q_4)=\mathcal{R}(c_2,q_4)=\mathcal{R}(c_{+},q_4)\ll 1$, whereas $\mathcal{R}_{-}(q_4)\equiv\mathcal{R}(c_3,q_4)=\mathcal{R}(c_4,q_4)=\mathcal{R}(c_{-},q_4)>0.143$, i.e., in the positive frequency domain at $|q_4-q_{40}|\ll 1$ the set of approximate solutions with the best accuracy has the two-dimensional amplitude subspace defined by the projection matrix $P_{+}$. By contrast, at $|q_4+q_{40}|\ll 1$ the projection matrix $P_{-}$ specifies the amplitude subspace, because in this negative frequency domain $\mathcal{R}_{-}(q_4)\ll 1$, but $\mathcal{R}_{+}(q_4)>0.143$. The graphic illustration of $\mathcal{R}_{-}(q_4)$ for $q_4<0$ can be obtained by the transformation $\mathcal{R}_{\mp}(q_4)=\mathcal{R}_{\pm}(-q_4)$.

Thus, for both $\Psi$~(\ref{psix}) and $\Psi'$~(\ref{psitr}) the amplitude subspace remain the same as for the free space solution. Nonetheless, sharp distinctions do exist. In free space, at any given $\textbf{q}$, there is the discrete spectrum of $q_4$ values, namely, $q_{40}$ and $-q_{40}$ (\ref{q4pm}). In the electromagnetic crystal under consideration, it is replaced by the continuous spectrum with two narrow domains in the vicinity of $\pm q_{40}$, which specify the family of approximate partial solutions with reasonable exactness. The rough initial approximation $\Psi'$ is sufficient to obtain the free space solution with its major features, the dispersion relation and the two-dimensional amplitude subspace, as the limiting case of vanishing field $I_A\rightarrow 0$. However, the function $\Psi$ provides a more accurate and detailed description of this limiting process (see Fig.~\ref{p3fig1}). The described above solutions domains with small values of $\mathcal{R}(c_{\pm},q_4)$ are very narrow and can be conveniently described in terms of the small variable
\begin{equation}\label{xiq4}
    \xi=q_4-q_{40}=\frac{\hbar \omega}{m_e c^2}-\sqrt{1+\left(\frac{\hbar \textbf{k}}{m_e c}\right)^2}
\end{equation}
at $q_4>0$ (see Fig.~\ref{p3fig1}) and $\xi=q_4+q_{40}$ at at $q_4<0$. For $\Psi$~(\ref{psix}) at $q_4>0$, the minimum value $\mathcal{R}_0$ of $\mathcal{R}_{+}$ and its position $\xi_0$ at the $\xi$-axis can be evaluated as
\begin{equation}\label{R0xi0}
    \mathcal{R}_0\approx 0.25A_m,\quad \xi_0\approx 0.5 I_A,
\end{equation}
where $I_A=12 A_m^2$ for ESTC1, and the width of the $\xi$--domain satisfying the condition $\mathcal{R}_{+}(\xi)\leq \sqrt{I_A}$ is approximately equal $I_A$.

\section{\label{sec:ESTC1}Spectral curve of approximate solutions}
Let us considerably enhance the amplitude $A_m$~(\ref{A16cr1}), in comparison to the values treated above, up to the value $A_m=5\times 10^{-4}$ ($I_A=3\times 10^{-6}$). In this case, it is necessary to use more elaborate finite $p$-models of ESTC1 described in~\cite{ESTCp2}. As before, we assume $\Omega=0.1, q_1=q_2=0, q_3=0.02$ and treat families of functions~\cite{ESTCp2}
\begin{equation}\label{psiSnc0}
    \Psi(\bm{x})=\sum_{n\in S_d}c(n)e^{i \varphi_n(\bm{x})}\equiv\sum_{n\in S_d}e^{i \varphi_n(\bm{x})}S(n)c_0
\end{equation}
with different values of $c_0=c_j, j=1,2\dots$, where $S_d$ is the solution domain, i.e., the subset of $\mathcal{L}$ with nonzero matrices $S(n)$. However, instead of the basis $c_j$ (\ref{c1234}), we use below the generalized eigenvectors $c_j$ defined by the equation
\begin{equation}\label{UDUE}
    U_Dc_j=\lambda_jU_Ec_j,
\end{equation}
where $U_E$ and $U_D$ are the Hermitian $4\times 4$ matrices which define the relative residual $\mathcal{R}$ as follows (see Eqs.~(26), (40) and (43) in \cite{ESTCp2})
\begin{equation}\label{Rc03}
   \mathcal{R}=\sqrt{\frac{c_0^{\dag} U_Dc_0}{c_0^{\dag} U_Ec_0}}.
\end{equation}
In the case under consideration, the quartic equation
\begin{equation}\label{quartic}
    \det(U_D -\lambda U_E)=0,
\end{equation}
which specifies the generalized eigenvalues $\lambda_j$, has real coefficients and twofold positive roots $\lambda_1$ and $\lambda_2$, indexed in increasing order of magnitude. The corresponding two--dimensional subspaces of generalized eigenvectors, i.e., bispinor amplitudes $c_1$ and $c_2$, are defined by the Hermitian projection matrices
\begin{equation}\label{roj}
    \rho_j=U-2\frac{U_D-\lambda_jU_E}{tr(U_D-\lambda_j U_E)},\quad j=1, 2.
\end{equation}
In the case that the amplitude $c_0$ satisfies the condition $\rho_jc_0=c_0$, Eq.~(\ref{Rc03}) gives $\mathcal{R}=\mathcal{R}_j\equiv\sqrt{\lambda_j}$. If $\mathcal{R}_j\ll 1$, the function $\Psi$ (\ref{psiSnc0}) with such amplitude is an approximate partial solution of the Dirac equation.

\begin{figure}
\includegraphics{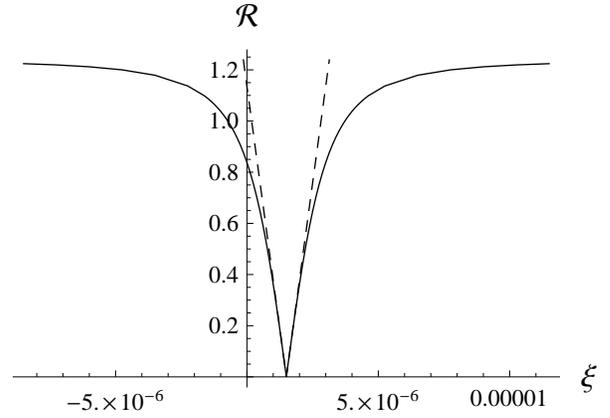}
\caption{\label{p3fig2}The functions $\mathcal{R}=\mathcal{R}_{1}(\xi)$ (solid curve) and $\mathcal{R}=\mathcal{R}_1^{ap}(\xi)$ (dash curve) for $1$-model of ESTC1 at  $A_m=5\times 10^{-4}$.}
\end{figure}

In the framework of any $p$-model of ESTC1, applied in this article ($p=0, 1, 2, 3$), the condition $\mathcal{R}_1\ll 1$ is satisfied within narrow limits of $\xi$ values (see Fig.~\ref{p3fig1}, Fig.~\ref{p3fig2}, and Table~\ref{tab:tab1}), whereas $\mathcal{R}_2$ does not satisfy the similar condition and $\mathcal{R}_2\gg\mathcal{R}_1$ at any value of $\xi$. The graphical representation of $\mathcal{R}=\mathcal{R}_1(\xi)$ will be denoted the spectral curve of approximate solutions. The minimum $\{\xi_0,\mathcal{R}_0=\mathcal{R}_1(\xi_0)\}$ of this curve specifies the most accurate approximate solution available in the frame of $p$-model under consideration.
\begin{table*}
\caption{\label{tab:tab1}Parameters of spectral curves for $p$-models of ESTC1.}
\begin{ruledtabular}
\begin{tabular}{lllll}
$p$&$\xi_0$&$\mathcal{R}_{1,2}$&$\beta_0$&$\delta\xi$\\ \hline
$0'$&$0$&$0.001732\,05$&$1$&$0$\\
    & &$2.00040$&   & \\
$0$&$1.492174147536518\times 10^{-6}$&$0.000123775$&$1152.74$&$1.49871\times 10^{-6}$\\
   & &$0.143302$& & \\
$1$&$1.499705741630043\times 10^{-6}$&$0.0000342977$&$757524$&$2.28602\times 10^{-9}$\\
   & &$2.02589$& & \\
$2$&$1.499679217218709\times 10^{-6}$&$1.00113\times 10^{-6}$&$4.15825\times 10^{7}$&$4.16534\times 10^{-11}$\\
   & &$0.902999$& & \\
$3$&$1.499679217930120\times 10^{-6}$&$1.72762\times 10^{-9}$&$2.95805\times 10^{10}$&$5.85538\times 10^{-14}$\\
   & &$2.04158$& & \\
\end{tabular}
\end{ruledtabular}
\end{table*}

%%%%%%%%%%%%%%%%%%%%%%%%%%%%%figure2%%%%%%%%%%%%%%%
%%%%%%%%%%%%%%%%%%%%%%%%%%%%%%%%%%%%%%%%%%%%%%%%%%

The bottom of curve $\mathcal{R}=\mathcal{R}_1(\xi)$, which is similar to the solid curves depicted in Fig.~\ref{p3fig1}, can be approximated as follows (see the dash curve in Fig.~\ref{p3fig2})
\begin{equation}\label{parabola}
    \mathcal{R}_1^{ap}(\xi)=\sqrt{\mathcal{R}_0^2+\beta_0^2(\xi-\xi_0)^2},
\end{equation}
where the values of $\xi_0, \mathcal{R}_0$ and $\beta_0$ for the $p$-models applied in this paper are presented in Table~\ref{tab:tab1} for the positive frequency domain $q_4>0$. This relation gives a rather close approximation of $\mathcal{R}_1(\xi)$, for example, in $1$-model illustrated in Fig.~\ref{p3fig2}, $\mathcal{R}_1^{ap}(\xi)/\mathcal{R}_1(\xi)-1~<0.012$ at the domain $5\times 10^{-8}<\xi<7\times 10^{-8}$, where $\mathcal{R}_1(\xi)<0.188$. Outside this bottom domain the dependence $\mathcal{R}_1$ on $\xi$ gradually becomes weak. Let $\mathcal{R}_{av}$ be an available (in $p$-model) level of the relative residual $\mathcal{R}$ and $\mathcal{R}_0,\mathcal{R}_{av}\ll 1$. The half-width $\delta\xi(\mathcal{R}_{av})$ of the solution line, i.e., the half-width of $\xi$ domain, where $\mathcal{R}_0\leq\mathcal{R}\leq\mathcal{R}_{av}$, can be estimated from Eq.~(\ref{parabola}) as
\begin{equation}\label{dxi}
    \delta\xi(\mathcal{R}_{av})=\frac{1}{\beta_0}\sqrt{\mathcal{R}_{av}^2-\mathcal{R}_0^2}.
\end{equation}
This half-width is a rapidly decreasing function of $p$. Table~\ref{tab:tab1} presents its values $\delta\xi=\delta\xi(\mathcal{R}_{av})$ at $\mathcal{R}_{av}=\sqrt{I_a}=\sqrt{3}\times 10^{-3}$.

\begin{figure}
\includegraphics{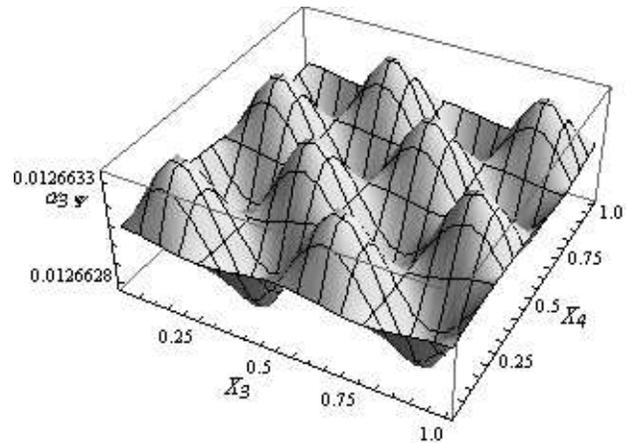}
\caption{\label{p3fig3}The Hermitian form $\alpha_{3\Psi}=\Psi^\dag(\bm{x})\alpha_3\Psi(\bm{x})$ at $X_1=X_2=0$ for ESTC1.}
\end{figure}

The projection matrix $\rho_1$ for ESTC1 is concisely defined by its Dirac set (see appendix in \cite{ESTCp1})
\begin{eqnarray}
% \nonumber to remove numbering (before each equation)
 D_s(\rho_1)=&&\{0.5, 0, 0, 0, 0.142825632163, 0, 0, 0, 0,\nonumber\\
 &&-0.479030979645,0.00406939990718,\nonumber\\
 &&0.00926170280712,-0.00527448410786,\nonumber\\
 &&0, 0, 0\}.
\end{eqnarray}
For any bispinor $c_a$, substituting $c_0=\rho_1c_a$ in (\ref{psiSnc0}) gives a partial solution with the same value of relative residual: $\mathcal{R}=\mathcal{R}_1$. In particular, one can use the free space basis $c_a=c_j$~(\ref{c1234}). This yields four different partial solutions with amplitudes $c_j^{(1)}=\rho_1c_j, j=1,2,3,4$. Of course, only two of them are linearly independent. To compare mean values of Hamiltonian $\langle{H}\rangle$, components of kinetic momentum  $\langle{p_k}\rangle$, probability current density $\langle{j_k}\rangle$,  and spin $\langle{S_k}\rangle$ for these solutions, we substitute $c_0=c_j^{(1)}$ and the operators
\begin{equation}\label{Hpk}
    H=c\sum_{k=1}^3\alpha_k p_k + m_e c^2\alpha_4,\quad p_k=-i\hbar\frac{\partial}{\partial x_k} - \frac{e}{c}A_k,
\end{equation}
\begin{equation}\label{jkSk}
    j_k=c\alpha_k, \quad S_k=\frac{\hbar}{2}\Sigma_k, \quad k=1, 2, 3,
\end{equation}
in Eqs.~(24)--(26) \cite{ESTCp2}. The calculations in the framework of 3-model result in the mean values
\begin{eqnarray}
% \nonumber to remove numbering (before each equation)
 \langle\mathcal{H}\rangle= \langle H\rangle /m_e c^2&=&1.000201480083165, \\
 \langle P_3\rangle=\langle p_3\rangle /m_e c&=&0.02000001996504673,\\
 \langle\alpha_3\rangle =\langle j_3\rangle/c&=&0.01999597119169098,
\end{eqnarray}
which are the same for these four solutions. However, the solutions have different mean values of spin components
\begin{eqnarray}
% \nonumber to remove numbering (before each equation)
  \langle\Sigma_1\rangle&=&\pm 6.4228848693\times 10^{-11} \text{ for } j=1, 2,\nonumber\\
   &=&\mp 0.017300503410 \text{ for } j=3, 4,\nonumber\\
  \langle\Sigma_2\rangle&=&\pm 1.4618089196\times 10^{-10} \text{ for } j=1, 2,\nonumber\\
    &=&\mp 0.038232859850\nonumber \text{ for } j=3, 4,\\
  \langle\Sigma_3\rangle&=&\pm 0.9999980004 \text{ for } j=1, 2,\nonumber\\
    &=&\pm 0.99911672861 \text{ for } j=3, 4.
\end{eqnarray}
Mean values of $p_{1,2}$ and $j_{1,2}$ are negligibly small: $|\langle{P_k}\rangle|, |\langle{\alpha_k}\rangle|< 10^{-20}$ for $k=1, 2$. Figure~\ref{p3fig3} illustrates the dependence of probability current density $j_3=c\alpha_{3\Psi}=c\Psi^\dag(\bm{x})\alpha_3\Psi(\bm{x})$ on the coordinates $X_3$ and $X_4$ at $X_1=X_2=0$.

It follows from the above numerical results that $\xi_0$ converges to a positive limit and $\mathcal{R}(\xi_0)$ tends to zero with increasing $p$, i.e., with expansion of a finite subsystem of equations described in~\cite{ESTCp2}. In the limit, $\Psi$~(\ref{psiSnc0}) converges to a family of exact solutions with the dispersion relation [see Eq.~(\ref{xiq4})]
\begin{equation}\label{dispeq}
    \frac{\hbar\omega}{m_e c^2}=\xi_0 + \sqrt{1+\left(\frac{\hbar\textbf{k}}{m_e c}\right)^2}
\end{equation}
and the two-dimensional amplitude subspace defined by $\rho_1=U-2U_D/ tr(U_D)$.

%%%%%%%%%%%%%%%%%%%%%%%%%%%%%figure3%%%%%%%%%%%%%%%
%%%%%%%%%%%%%%%%%%%%%%%%%%%%%%%%%%%%%%%%%%%%%%%%%%%

\section{\label{sec:ESTC2}ESTC composed of circularly polarized waves: spin birefringence}
In this section, we treat ESTC2 composed of six circularly polarized waves with the amplitudes (see Eqs.~(2) and (4) in~\cite{ESTCp1})
\begin{eqnarray}\label{Acr2}
    \textbf{A}_1&=&\textbf{A}_4=A_m(\textbf{e}_2+i \textbf{e}_3)/\sqrt{2},\nonumber\\
    \textbf{A}_2&=&\textbf{A}_5=A_m(\textbf{e}_3+i \textbf{e}_1)/\sqrt{2},\nonumber\\
    \textbf{A}_3&=&\textbf{A}_6=A_m(\textbf{e}_1+i \textbf{e}_2)/\sqrt{2},
\end{eqnarray}
where $A_m=5\times 10^{-4}$ and $I_A=3\times 10^{-6}$ take the same values as in the case of EmCr1 treated above. The parameters  $\Omega=0.1, q_1=q_2=0$ and $q_3=0.02$ are also retain their previous values, so that we change only the polarizations of electromagnetic waves from linear to circular. To study the properties of ESTC2, we apply $p$-models with $p=0, 1, 2, 3$.

In the case of ESTC2, Eq.~(\ref{quartic}) has four different real positive roots: $\lambda_1<\lambda_2<\lambda_3<\lambda_4$, and the generalized eigenvectors $c_j$ satisfy the orthogonality relations
\begin{equation}\label{ciortho}
    c_i^{\dag}U_E c_j=0,\; c_i^{\dag}U_D c_j=0,\; i\neq j,\; i,j=1,2,3,4.
\end{equation}
The corresponding generalized one-dimensional eigen subspaces are uniquely defined by the Hermitian projection matrices (dyads)
\begin{equation}\label{dyad}
    \rho_j = \frac{c_j\otimes c_j^{\dag}}{c_j^{\dag}c_j} = \frac{\overline{D_j}}{tr(\overline{ D_j})},\quad j=1,2,3,4,
\end{equation}
where $D_j=U_D - \lambda_j U_E$, $\overline{D}$ is the adjoint matrix, $D\overline{D}=\overline{D}D=|D|U$. In the limiting case, when $\mathcal{R}(\xi_0)$ tends to zero with increasing $p$, $\rho_1=\overline{U_D}/tr(\overline{U_D})$. It is significant that these matrices are the uniquely defined descriptors of the subspaces in contrast to basis elements $c_j$. This provides a convenient means to use real Dirac sets of Hermitian projection matrices for comparative analysis of subspaces.

\begin{figure}
\includegraphics{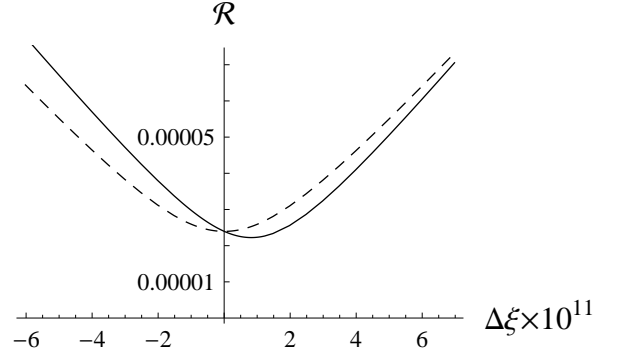}
\caption{\label{p3fig4}The dependence of $\mathcal{R}$ on $\Delta\xi=\xi-\xi_{0a}$ in $1$-model of ESTC2: the dash line (line~$a$) is defined by $\mathcal{R}_1=\sqrt{\lambda_1}$ at $\Delta\xi\leq 0$ and by $\mathcal{R}_2=\sqrt{\lambda_2}$ at $\Delta\xi>0$; the solid line (line~$b$) is defined by $\mathcal{R}_2$ at $\Delta\xi\leq 0$ and by $\mathcal{R}_1$ at $\Delta\xi>0$.}
\end{figure}

The solution curve $\mathcal{R}=\mathcal{R}(\xi)$ in ESTC2 splits into two doublet lines called below line~$a$ and line~$b$ with minimum at $\xi_0=\xi_{0a}$ and $\xi_0=\xi_{0b}$, respectively, see Table~\ref{tab:tab2} and Fig.~\ref{p3fig4}. Although the doublet lines are very close, D-sets of $\rho_{1a}=\rho_1(\xi_{0a})$ and $\rho_{1b}=\rho_1(\xi_{0b})$~(\ref{dyad}) considerably differ from one another, in $3$-model they are given by
\begin{eqnarray*}\label{ro1ab}
% \nonumber to remove numbering (before each equation)
  D_s(\rho_{1a})=&&\left\{0.25, 0.206495, -0.00688536,\right.\\
   &&-0.00688574, -0.102352, -0.117887,\\
   &&0.0903974, 0.0903979, 0.179608,\\
   &&-0.220245, -0.0419267, -0.0419338,\\
   &&-6.70254\times 10^{-6}, 8.04274\times10^{-6},\\
   &&\left. 0.0994123, -0.0994089\right\}, \\
  D_s(\rho_{1b})=&&\left\{0.25, 0.125694, 0.0236921,\right.\\
   &&0.023691, -0.119579, -0.200412,\\
   &&-0.0992794, -0.0992784, 0.0512099,\\
   &&-0.145645, 0.116162, 0.116167, \\
   && -3.80381\times 10^{-7}, 2.68718\times 10^{-6},\\
   &&\left.-0.150963, 0.150959\right\}.
\end{eqnarray*}

The finite $p$-models of ESTC2 with $p=0, 1$ are qualified mainly for fast scanning of $\mathcal{R}(\xi)$ in preliminary search of minimums. This is necessary because in the frame of $p$-models with $p=2, 3$, which provide approximate solutions with much better accuracy, the solution domain width  $\delta\xi$ becomes very small. Most important of all, $\mathcal{R}_1$ rapidly decreases whereas $\mathcal{R}_2$ increases with increasing $p$ for the both doublet lines, see Table~\ref{tab:tab2}. Because of this, only line~$a$  provides the solution at $\xi=\xi_{0a}$, whereas line~$b$ provides the solution at $\xi=\xi_{0b}$.

\begin{figure}
\includegraphics{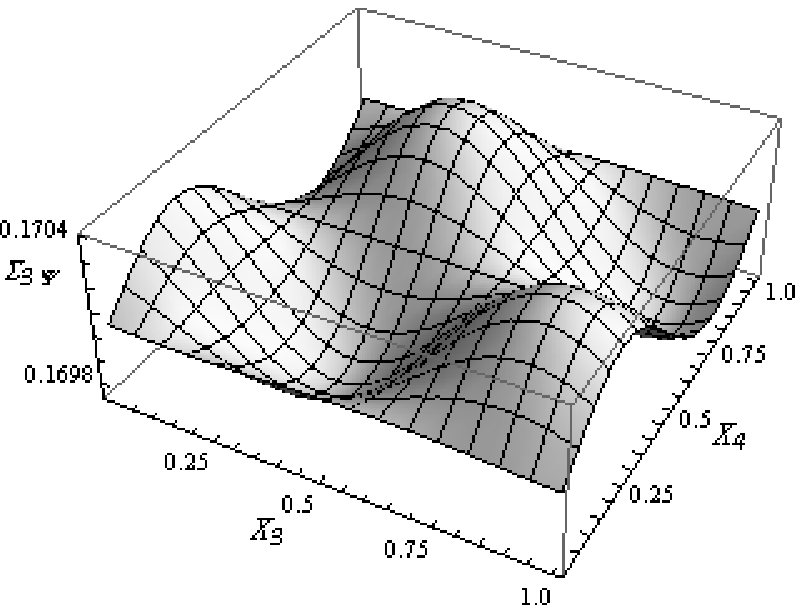}
\caption{\label{p3fig5}The Hermitian form $\Sigma_{3\Psi}=\Psi^\dag(\bm{x})\Sigma_3\Psi(\bm{x})$ at $X_1=X_2=0$ for line~$a$ in ESTC2.}
\end{figure}

\begin{figure}
\includegraphics{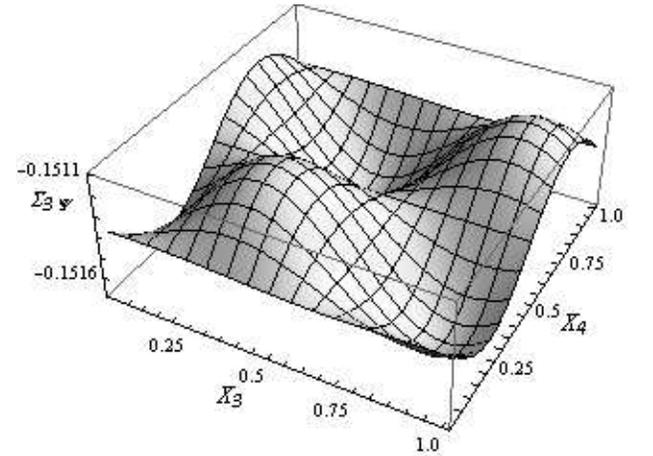}
\caption{\label{p3fig6}The Hermitian form $\Sigma_{3\Psi}=\Psi^\dag(\bm{x})\Sigma_3\Psi(\bm{x})$ at $X_1=X_2=0$ for line~$b$ in ESTC2.}
\end{figure}

Table~\ref{tab:tab3} presents mean values of operators $H$, $p_k$, $\alpha_k$ and $\Sigma_k$ with respect to the functions $\Psi$~(\ref{psiSnc0}) with the amplitude $c_0$ satisfying the conditions $\rho_{1a}c_0=c_0$ and $\rho_{1b}c_0=c_0$ at $\xi=\xi_{0a}$ and $\xi=\xi_{0b}$, respectively, calculated for $3$-model of ESTC2. The major difference between these two partial solutions for doublet lines manifests itself in spin projections (see Table~\ref{tab:tab3} and Figs.~\ref{p3fig5} and~\ref{p3fig6}). In other words, the electromagnetic crystal formed by circularly polarized waves possesses the spin birefringence. It reveals itself as the splitting of Eq.~(\ref{dispeq}) into two dispersion relations with $\xi_0=\xi_{0a}$ and $\xi_0=\xi_{0b}$. For a given wave vector $\textbf{k}$ they provide frequencies $\omega_a$ and $\omega_b$, which specify two different partial solutions $\Psi$~(\ref{psiSnc0}) with one-dimensional amplitude subspaces defined by $\rho_{1a}$ and $\rho_{1b}$.

%%%%%%%%%%%%%%%%%%%%%%%%%%%%%figure4%%%%%%%%%%%%%%%
%%%%%%%%%%%%%%%%%%%%%%%%%%%%%%%%%%%%%%%%%%%%%%%%%%

%%%%%%%%%%%%%%%%%%%%%%table2%%%%%%%%%%%%%%%%%%%%%%
\begin{table*}
\caption{\label{tab:tab2}Parameters of spectral curves for $p$-models of ESTC2.}
\begin{ruledtabular}
\begin{tabular}{lllll}
$p$&$\xi_{0a},\xi_{0b}, (\xi_{0b} -\xi_{0a})$&$\mathcal{R}_{1,2,3,4}$&$\beta_0$&$\delta\xi$\\ \hline
$0$&$1.451\,475\,281\,655\,971\times 10^{-6}$&$0.000\,110\,130$&$1185.08$&$1.458\,59\times 10^{-6}$\\
   & &$0.000\,151\,367$& & \\
   & &$0.143\,298$& & \\
   & &$0.143\,298$& & \\
$ $&$1.537\,787\,434\,029\,065\times 10^{-6}$&$0.000\,116\,519$&$1119.42$&$1.543\,77\times 10^{-6}$\\
   &($8.631\,215\,237\times 10^{-8}$)&$0.000\,150\,303$& & \\
   & &$0.143\,298$& & \\
   & &$0.143\,298$& & \\
$1$&$1.499\,696\,566\,656\,439\times 10^{-6}$&$0.000\,023\,9446$&$1.034\,61\times 10^6$&$1.673\,95\times 10^{-9}$\\
   & &$0.000\,023\,9682$& & \\
   & &$2.007\,66$& & \\
   & &$2.010\,59$& & \\
$ $&$1.499\,704\,784\,068\,618\times 10^{-6}$&$0.000\,022\,2381$&$1.038\,25\times 10^6$&$1.668\,11\times 10^{-9}$\\
   &($8.217\,412\,179\times 10^{-12}$)&$0.000\,025\,2824$& & \\
   & &$2.007\,66$& & \\
   & &$2.010\,59$& & \\
   $2$&$1.499\,676\,893\,270\,552\times 10^{-6}$&$1.217\,02\times 10^{-6}$&$3.803\,69\times 10^7$&$4.553\,61\times 10^{-11}$\\
   & &$0.000\,196\,542$& & \\
   & &$1.993\,64$& & \\
   & &$2.009\,47$& & \\
$ $&$1.499\,681\,284\,944\,732\times 10^{-6}$&$1.048\,17\times 10^{-6}$&$4.474\,26\times 10^7$&$3.871\,14\times 10^{-11}$\\
   &($4.391\,674\,179\times 10^{-12}$)&$0.000\,167\,009$& & \\
   & &$1.993\,64$& & \\
   & &$2.009\,47$& & \\
 $3$&$1.499\,676\,894\,704\,856\times 10^{-6}$&$2.174\,76\times 10^{-9}$&$2.321\,00\times 10^{10}$&$7.459\,32\times 10^{-14}$\\
   & &$0.106\,985$& & \\
   & &$1.911\,97$& & \\
   & &$2.057\,48$& & \\
$ $&$1.499\,681\,285\,335\,523\times 10^{-6}$&$1.816\,45\times 10^{-9}$&$2.439\,65\times 10^{10}$&$7.099\,58\times 10^{-14}$\\
   &($4.390\,630\,667\times 10^{-12}$)&$0.101\,811$& & \\
   & &$1.911\,97$& & \\
   & &$2.057\,48$& & \\
\end{tabular}
\end{ruledtabular}
\end{table*}
%%%%%%%%%%%%%%%%%%%%%%%%%%%%%%%%%%%%%%%%%%

%%%%%%%%%%%%%%%%%%%%%%table3%%%%%%%%%%%%%%%%%%%%%%
\begin{table}
\caption{\label{tab:tab3}Mean values $\langle\mathcal{H}\rangle$, $\langle{P_k}\rangle$, $\langle{\alpha_k}\rangle$ and $\langle{\Sigma_k}\rangle, k=1,2,3$ for $3$-model of ESTC2.}
\begin{ruledtabular}
\begin{tabular}{lll}
quantity & \text{line}~$a$&\text{line}~$b$\\ \hline
$\langle\mathcal{H}\rangle$&$1.000201393528605$&$1.000201566634309$\\
$\langle{P_1}\rangle=\langle{P_2}\rangle$&$5.9364197888\times 10^{-10}$&$-5.9256695546\times 10^{-10}$\\
$\langle{P_3}\rangle$&$0.0200000193646438$&$0.0200000205583700$\\
$\langle{\alpha_1}\rangle=\langle{\alpha_2}\rangle$&$5.32701283\times 10^{-13}$&$5.6468670\times 10^{-13}$\\
$\langle{\alpha_3}\rangle$&$0.0199959711849530$&$0.01999597119136176$\\
$\langle{\Sigma_1}\rangle$&$0.568943120867496$&$-0.568943120731802$\\
$\langle{\Sigma_2}\rangle$&$0.568943120867590$&$-0.568943120731907$\\
$\langle{\Sigma_3}\rangle$&$0.593586130698993$&$-0.593586131019241$\\
\end{tabular}
\end{ruledtabular}
\end{table}

%%%%%%%%%%%%%%%%%%%%%%%%%%%%%figure5%%%%%%%%%%%%%%%
%%%%%%%%%%%%%%%%%%%%%%%%%%%%%%%%%%%%%%%%%%%%%%%%%%

%%%%%%%%%%%%%%%%%%%%%%%%%%%%%figure6%%%%%%%%%%%%%%%
%%%%%%%%%%%%%%%%%%%%%%%%%%%%%%%%%%%%%%%%%%%%%%%%%%

\section{Concluding remarks}
The electromagnetic crystals is a family of periodic fields specified by complex vector amplitudes of six plane harmonic waves forming a crystal and the field frequency. These crystals have a specific impact on the motion of electrons, which may result in such interesting effects as spin birefringence. In this paper, we have restricted our consideration to the particular case with the fixed value of the wave vector $\textbf{k}$ in Eq.~(\ref{phin}). Results of an investigation into the dependence of ESTCs properties on the magnitude and the direction of $\textbf{k}$, in particular, the energy band structure of ESTCs, will be discussed in our subsequent papers.

The fundamental solution of the Dirac equation and the techniques presented in this series of papers provide a means for detailed study of the electron motion in ESTCs. Some of these techniques, in particular, the method~\cite{ESTCp1} for calculating the fundamental solution of a system of homogeneous linear equations, the fractal approach~\cite{ESTCp2} to expansion of subsystems of equations at calculating approximate solutions, and the use of the relative residual $\mathcal{R}$~\cite{ESTCp2} at the comparative analysis of families of approximate solutions, may be also useful in solving other problems in mathematical and theoretical physics.

\appendix*
\section{}
As we have shown in Ref.~\cite{ESTCp1}, the Dirac equation describing the motion of  an electron in ESTC reduces to to the infinite system of linear equations relating Fourier amplitudes [bispinors $c(n)$] of the wave function $\Psi$. The interconnections of equations depend on complex vector amplitudes $\bm{A}_j, j=1,2,\dots,6$ of six plane waves forming ESTC, for example, see Eqs.~(\ref{A16cr1}) and (\ref{Acr2}). These interconnections are described by 12 matrix functions  $N_1(m,s)$ with $m, s\in \mathcal{L}, g_{4d}(s)=1$ and 56 scalar coefficients $N_2(s)$ with $g_{4d}(s)=2$. The definitions of $N_1(m,s)$ and $N_2(s)$ are given in Ref.~\cite{ESTCp1}. Here, we present these major structural parameters in the explicit form that is necessary in any numerical implementation of the general techniques developed in Refs.~\cite{ESTCp1,ESTCp2}.

\subsection{Dirac sets of matrices $N_1(m,s)$}
We present $N_1(m,s)$ and $N_2(s)$ in order of the sequential numbering $i=0,1,\dots$ of points $s=s(i)\in \mathcal{L}$ (see appendix in Ref.~\cite{ESTCp2}). Let $A_{jk}$ be the Cartesian components of $\bm{A}_j$,  $m=(m_1,m_2,m_3,m_4)\in \mathcal{L}$, $w_k=q_k+m_k\Omega$ and $\Omega_{\pm}=\pm\Omega+2w_4$. There are 12 points with $g_{4d}(s)=1$. They are elements (from 2 to 13) of the list $S_{69}$ \cite{ESTCp2}. The Dirac sets (see appendix in~\cite{ESTCp1}) of matrices $N_1[m,s(i)], i=1,\dots,12$ have the form:
\begin{eqnarray*}
% \nonumber to remove numbering (before each equation)
  D_s&&\left\{N_1[m,(0,0,-1,-1)]\right\}=\\
     &&\left\{-2(A_{31}w_1+A_{32}w_2), 0,i A_{32}\Omega, -i A_{31}\Omega,0,0,0,0,\right.\\
     &&\left. 0,0, -A_{31}\Omega_{-},-A_{32}\Omega_{-},0,0,0,0\right\},
\end{eqnarray*}
\begin{eqnarray*}
% \nonumber to remove numbering (before each equation)
  D_s&&\left\{N_1[m,(0,-1,0,-1)]\right\}=\\
     &&\left\{-2(A_{21}w_1+A_{23}w_3),i A_{21}\Omega,-i A_{23}\Omega,0,0,0,0,0,\right.\\
     &&\left. 0, -A_{23}\Omega_{-},-A_{21}\Omega_{-},0,0,0,0,0\right\},
\end{eqnarray*}
\begin{eqnarray*}
% \nonumber to remove numbering (before each equation)
  D_s&&\left\{N_1[m,(-1,0,0,-1)]\right\}=\\
     &&\left\{-2(A_{12}w_2+A_{13}w_3),-i A_{12}\Omega,0,i A_{13}\Omega,0,0,0,0,\right.\\
     &&\left. 0, -A_{13}\Omega_{-},0,-A_{12}\Omega_{-},0,0,0,0\right\},
\end{eqnarray*}
\begin{eqnarray*}
% \nonumber to remove numbering (before each equation)
  D_s&&\left\{N_1[m,(1,0,0,-1)]\right\}=\\
     &&\left\{-2(A_{42}w_2+A_{43}w_3),i A_{42}\Omega,0,-i A_{43}\Omega,0,0,0,0,\right.\\
     &&\left. 0, -A_{43}\Omega_{-},0,-A_{42}\Omega_{-},0,0,0,0\right\},
\end{eqnarray*}
\begin{eqnarray*}
% \nonumber to remove numbering (before each equation)
  D_s&&\left\{N_1[m,(0,1,0,-1)]\right\}=\\
     &&\left\{-2(A_{51}w_1+A_{53}w_3),-i A_{51}\Omega,i A_{53}\Omega,0,0,0,0,0,\right.\\
     &&\left. 0, -A_{53}\Omega_{-},-A_{51}\Omega_{-},0,0,0,0,0\right\},
\end{eqnarray*}
\begin{eqnarray*}
% \nonumber to remove numbering (before each equation)
  D_s&&\left\{N_1[m,(0,0,1,-1)]\right\}=\\
     &&\left\{-2(A_{61}w_1+A_{62}w_2),0,-i A_{62}\Omega,i A_{61}\Omega,0,0,0,0,\right.\\
     &&\left. 0,0,-A_{61}\Omega_{-},-A_{62}\Omega_{-},0,0,0,0\right\},
\end{eqnarray*}
\begin{eqnarray*}
% \nonumber to remove numbering (before each equation)
  D_s&&\left\{N_1[m,(0,0,-1,1)]\right\}=\\
     &&\left\{-2(A^{\ast}_{61}w_1+A^{\ast}_{62}w_2),0,i A^{\ast}_{62}\Omega,-i A^{\ast}_{61}\Omega,0,0,0,0,\right.\\
     &&\left. 0,0, -A^{\ast}_{61} \Omega_{+},-A^{\ast}_{62}\Omega_{+},0,0,0,0\right\},
\end{eqnarray*}
\begin{eqnarray*}
% \nonumber to remove numbering (before each equation)
  D_s&&\left\{N_1[m,(0,-1,0,1)]\right\}=\\
     &&\left\{-2(A^{\ast}_{51}w_1+A^{\ast}_{53}w_3),i A^{\ast}_{51}\Omega,-i A^{\ast}_{53}\Omega,0,0,0,0,0,\right.\\
     &&\left. 0, -A^{\ast}_{53}\Omega_{+}, -A^{\ast}_{51}\Omega_{+},0,0,0,0,0\right\},
\end{eqnarray*}
\begin{eqnarray*}
% \nonumber to remove numbering (before each equation)
  D_s&&\left\{N_1[m,(-1,0,0,1)]\right\}=\\
     &&\left\{-2(A^{\ast}_{42}w_2+A^{\ast}_{43}w_3),-i A^{\ast}_{42}\Omega,0,i A^{\ast}_{43}\Omega,0,0,0,0,\right.\\
     &&\left. 0,-A^{\ast}_{43}\Omega_{+},0, -A^{\ast}_{42}\Omega_{+},0,0,0,0\right\},
\end{eqnarray*}
\begin{eqnarray*}
% \nonumber to remove numbering (before each equation)
  D_s&&\left\{N_1[m,(1,0,0,1)]\right\}=\\
     &&\left\{-2(A^{\ast}_{12}w_2+A^{\ast}_{13}w_3),i A^{\ast}_{12}\Omega,0,-i A^{\ast}_{13}\Omega,0,0,0,0,\right.\\
     &&\left. 0,-A^{\ast}_{13}\Omega_{+},0, -A^{\ast}_{12}\Omega_{+},0,0,0,0\right\},
\end{eqnarray*}
\begin{eqnarray*}
% \nonumber to remove numbering (before each equation)
  D_s&&\left\{N_1[m,(0,1,0,1)]\right\}=\\
     &&\left\{-2(A^{\ast}_{21}w_1+A^{\ast}_{23}w_3),-i A^{\ast}_{21}\Omega,i A^{\ast}_{23}\Omega,0,0,0,0,0,\right.\\
     &&\left. 0,-A^{\ast}_{23}\Omega_{+}, -A^{\ast}_{21}\Omega_{+},0,0,0,0,0\right\},
\end{eqnarray*}
\begin{eqnarray*}
% \nonumber to remove numbering (before each equation)
  D_s&&\left\{N_1[m,(0,0,1,1)]\right\}=\\
     &&\left\{-2(A^{\ast}_{31}w_1+A^{\ast}_{32}w_2),0,-i A^{\ast}_{32}\Omega,i A^{\ast}_{31}\Omega,0,0,0,0,\right.\\
     &&\left. 0,0,-A^{\ast}_{31}\Omega_{+}, -A^{\ast}_{32}\Omega_{+},0,0,0,0\right\}.
\end{eqnarray*}

\subsection{Coefficients $N_2(s)$ }
There are 56 points $s=s(i)\in \mathcal{L}, i=13,\dots,68$ with $g_{4d}(s)=2$. They are elements (from 14 to 69) of the list $S_{69}$ \cite{ESTCp2}. The list of the coefficients $N_2(s)$ has the form
%\begin{widetext}
\begin{eqnarray*}
% \nonumber to remove numbering (before each equation)
   \{N_2[s(i)],i=13,\dots,68\} =\\
   \left\{2 \left(A_{12} A_{42} + A_{13} A_{43} + A_{21} A_{51} + A_{23} A_{53}\right.\right.\\
   \left. + A_{31} A_{61} + A_{32} A_{62}), \right.\\
   2 \left(A^{\ast}_{12} A^{\ast}_{42} + A^{\ast}_{13} A^{\ast}_{43} + A^{\ast}_{21} A^{\ast}_{51} + A^{\ast}_{23} A^{\ast}_{53}\right.\\
   \left. + A^{\ast}_{31} A^{\ast}_{61} + A^{\ast}_{32} A^{\ast}_{62}\right),\\
   2 (A_{31} A^{\ast}_{61} + A_{32} A^{\ast}_{62}), 2 (A_{31} A^{\ast}_{51} + A_{21} A^{\ast}_{61}),\\
   2 (A_{32} A^{\ast}_{42} + A_{12} A^{\ast}_{62}), 2 (A^{\ast}_{12} A_{32} + A_{42} A^{\ast}_{62}),\\
   2 (A^{\ast}_{21} A_{31} + A_{51} A^{\ast}_{61}), 2 (A_{21} A^{\ast}_{51} + A_{23} A^{\ast}_{53}),\\
   2 (A_{23} A^{\ast}_{43} + A_{13} A^{\ast}_{53}), 2 (A^{\ast}_{13} A_{23} + A_{43} A^{\ast}_{53}),\\
   2 (A_{12} A^{\ast}_{42} + A_{13} A^{\ast}_{43}), 2 (A^{\ast}_{12} A_{42} + A^{\ast}_{13} A_{43}),\\
   2 (A_{13} A^{\ast}_{23} + A^{\ast}_{43} A_{53}), 2 (A^{\ast}_{23} A_{43} + A^{\ast}_{13} A_{53}),\\
   2 (A^{\ast}_{21} A_{51} + A^{\ast}_{23} A_{53}), 2 (A_{21} A^{\ast}_{31} + A^{\ast}_{51} A_{61}),\\
   2 (A_{12} A^{\ast}_{32} + A^{\ast}_{42} A_{62}), 2 (A^{\ast}_{32} A_{42} + A^{\ast}_{12} A_{62}),\\
   2 (A^{\ast}_{31} A_{51} + A^{\ast}_{21} A_{61}), 2 (A^{\ast}_{31} A_{61} + A^{\ast}_{32} A_{62}),\\
   (A_{31} + i A_{32}) (A_{31} - i A_{32}), 2 A_{21} A_{31}, 2 A_{12} A_{32},\\
   2 A_{32} A_{42}, 2 A_{31} A_{51}, (A_{21} + i A_{23}) (A_{21} - i A_{23}),\\
   2 A_{13} A_{23}, 2 A_{23} A_{43}, (A_{12} + i A_{13}) (A_{12} - i A_{13}),\\
   (A_{42} + i A_{43}) (A_{42} - i A_{43}), 2 A_{13} A_{53}, 2 A_{43} A_{53},\\
   (A_{51} + i A_{53}) (A_{51} - i A_{53}), 2 A_{21} A_{61}, 2 A_{12 }A_{62},\\
   2 A_{42} A_{62}, 2 A_{51} A_{61}, (A_{61} + i A_{62}) (A_{61} - i A_{62}),\\
   (A^{\ast}_{61} + i A^{\ast}_{62}) (A^{\ast}_{61} - i A^{\ast}_{62}), 2 A^{\ast}_{51} A^{\ast}_{61}, 2 A^{\ast}_{42} A^{\ast}_{62},\\
   2A^{\ast}_{12} A^{\ast}_{62}, 2 A^{\ast}_{21} A^{\ast}_{61}, (A^{\ast}_{51} + i A^{\ast}_{53}) (A^{\ast}_{51} - i A^{\ast}_{53}),\\
   2 A^{\ast}_{43} A^{\ast}_{53}, 2 A^{\ast}_{13} A^{\ast}_{53}, (A^{\ast}_{42} + i A^{\ast}_{43}) (A^{\ast}_{42} - i A^{\ast}_{43}),\\
   (A^{\ast}_{12} + i A^{\ast}_{13}) (A^{\ast}_{12} - i A^{\ast}_{13}), 2 A^{\ast}_{23} A^{\ast}_{43}, 2 A^{\ast}_{13} A^{\ast}_{23},\\
   (A^{\ast}_{21} + i A^{\ast}_{23}) (A^{\ast}_{21} - i A^{\ast}_{23}), 2 A^{\ast}_{31} A^{\ast}_{51}, 2 A^{\ast}_{32} A^{\ast}_{42},\\
   \left.  2 A^{\ast}_{12} A^{\ast}_{32}, 2 A^{\ast}_{21} A^{\ast}_{31}, (A^{\ast}_{31} + i A^{\ast}_{32}) (A^{\ast}_{31} - i A^{\ast}_{32})\right\}.
\end{eqnarray*}
\bibliography{Borzdov1}
\end{document}